\documentclass[11pt,a4paper]{article}
\usepackage{jheppub}
\usepackage{color,subfigure}

\begin{document}

\title{Resonances of Kalb-Ramond field on symmetric
       and asymmetric thick branes}
\author{Yun-Zhi Du,
        Li Zhao\footnote{Corresponding author},
        Yi Zhong,
        Chun-E Fu,
        Heng Guo}
\affiliation[]{Institute of Theoretical Physics,
    Lanzhou University, Lanzhou 730000,
    People's Republic of China}
\emailAdd{duyz11@lzu.edu.cn}
\emailAdd{lizhao@lzu.edu.cn}
\emailAdd{zhongy11@lzu.edu.cn}
\emailAdd{fuche08@lzu.edu.cn}
\emailAdd{guoh06@lzu.edu.cn}

\abstract{In this paper, we investigate the localization of the Kalb-Ramond field on symmetric and asymmetric thick branes, which are generated by a background scalar field. In order to localize the Kalb-Ramond field, we introduce a coupling with the background scalar field, and find that there exist some Kaluza-Klein resonant modes. For the case of symmetric brane, we seek the resonances by using the relative probability method and transfer matrix method, and obtain the same result for the two methods. For the asymmetric case, we use the transfer matrix method. We find that the number of resonances will decrease with the increase of the asymmetry.
}


\maketitle

\section{Introduction}\label{scheme1}

In 1920s, in order to unify electromagnetism with Einstein Gravity, Kaluza and Klein (KK) considered a five-dimensional space-time with one extra dimension \cite{1921}. And in the string/M-theory, it is required that there are six/seven extra dimensions. But in these theories, the size of extra dimensions is restricted to be at the Planck length, which is too small to be detected by the present experiments. Subsequently, the braneworld scenario was introduced in a higher-dimensional space-time. The representative brane models are the Arkani-Hamed-Dimopoulos-Dvali (ADD) model \cite{Arkani-Hamed1998,Arkani-Hamed1998a,Antoniadis1998} and the Randall-Sundrum (RS) one \cite{Randall1999,Randall1999a}. In the brane models, the particles in Standard Model are confined on the brane, while gravity can propagate in the whole space-time.

In the framework of braneworld scenario, extra dimensions do not have to be so small as the Planck length; they even can be infinite. So it is possible to detect the extra dimensions. And the localization of matters on a brane is a way to realize it. Because the matters from the higher dimensional space-time carry the information of extra dimensions, so if they can be localized on the brane, there is a hope for us to find them, and so prove the existence of extra dimensions. The localization of various matters has been investigated in the thin brane models \cite{George2007,Zhang2008,Guerrero2010,Liu2012a,Yang2012} and thick brane models \cite{Gremm2000,Csaki2000,Liu2008,Bazeia2009,Almeida2009,Liu2009,Liu2010,Zhao2010,Zhao2011,Fu2011,Li2011,Guo2011,Guo2012,Guo2012a,Fu2012,Liu2012b,Gogberashvili2012,HoffdaSilva2012}. It was found that scalar and gravity fields can be localized on the RS-like branes in the five-dimensional space-time with one extra dimension, but it is not true for vector and Kalb-Ramond (KR) tensor fields \cite{Neronov2001,Cruz2009,Tahim2009,Christiansen2012}. However, vector can be localized on some de sitter thick branes \cite{Liu2009} and Weyl thick branes \cite{Liu2008a}.

The KR field is an anti-symmetric tensor field, which was first introduced in the string theory. It is known that in the four-dimensional space-time, the antisymmetric tensor fields are equivalent to scalar or vector fields by the symmetry known as duality \cite{Quevedo2010}. But they will indicate new types of particles in extra dimensions. So we are interested in this field. However, the free KR field cannot be localized on the brane in five-dimensional space-time \cite{Cruz2009}. In Refs. \cite{Tahim2009,Fu2011,Christiansen2012}, the authors showed that, if the KR field couples to a dilaton field, then the localization can be realized. While in this paper, we find if we introduce the coupling with the background kink scalar, the KR field can also be localized on the brane under some conditions.

Here we consider an interesting brane model, i.e., the symmetric and asymmetric branes, which was investigated in Refs. \cite{Melfo2003,Guerrero2005,Liu2009,Liu2009a}. In this model, there is an asymmetric factor $a$ describing the asymmetry of the brane, and the brane can be split. The localization of fermion fields in this model has been investigated in \cite{Liu2009,Liu2009a}, where the authors mainly focused on the symmetric case. With the relative probability method applied in \cite{Liu2009,Liu2009a,Liu2009c}, the fermion resonances were found for the symmetric case. However, it is seem that this method is not easy to be extended to the asymmetric case.

In this work, we will use both the relative probability method and the transfer matrix method to investigate the KR massive KK modes. The transfer matrix method was used in Refs. \cite{Landim2011,Landim2012,Alencar2012} in the framework of braneworld, where the authors defined a transmission coefficient $T$ to find the resonances. And we will use the method to seek the resonances of the KR field on both symmetric and asymmetric branes, and study the effect of the asymmetry of the brane on the resonances. For the symmetric case, we compare this method with the relative probability method, and obtain the same result.

This paper is organized as follows. Firstly, we simply give in the next section a review of the construction of symmetric and asymmetric thick branes in five-dimensional space-time. Then we investigate the resonances of the KR field on the symmetric case in Sec.~\ref{scheme2}, and on the asymmetric case in Sec.~\ref{sec22}. Finally, the discussion and conclusion are given in Sec.~\ref{secConclusion}.

\section{Kalb-Ramond field on symmetric and asymmetric thick branes}{\label{sec2}}
\label{review}

We consider the thick brane generated by a single real scalar field $\phi$ in a five-dimensional space-time. The action of this system is
\begin{equation}
S=\int d^5x\sqrt{-g}~
   \Big[
         \frac{1}{2\kappa^2_5}R
         -\frac{1}{2}g^{MN}\partial_M
           \phi\partial_N\phi
         -U(\phi)
   \Big],\label{S}
\end{equation}
where $\kappa^2_5=8\pi G_5$ with $G_5$ the Newton Gravitational constant, and $U(\phi)$ is the scalar potential. In this paper we set $\kappa_5=1$ for simplicity.

The line-element of the space-time is assumed as
\begin{equation}
ds^2=e^{2A(z)}(\eta_{\mu\nu}dx^\mu dx^\nu+dz^2).
\label{metric}
\end{equation}
Here $z$ stands for the extra coordinate, and $e^{2A(z)}$ is the warp factor. The warp factor and the scalar field are both supposed to be functions of $z$ only, i.e., $\phi=\phi(z)$, $A=A(z)$.

A solution for this system was found in Ref. \cite{Guerrero2002}:
\begin{eqnarray}
\phi&=&\phi_0\arctan(\lambda z)^s,\\
e^{2A}&=&\frac{1}{[1+(\lambda z)^{2s}]^{1/s}G(z)^2},\label{warpfactor}\\
U(\phi)&=&-\frac{3}{4}\sin^2(\phi/\phi_0)
            \tan^{-2/s}(\phi/\phi_0)
             {\cal K}(\phi)\times
              \Big\{
                     16a\tan^{1/s}(\phi/\phi_0)
                      +\cos^{-2/s}(\phi/\phi_0)\nonumber\\
                    &&\big[
                            5-2s-(3+2s)\cos(2\phi/\phi_0)
                      \big]{\cal K}(\phi)
              \Big\}
              -6a^2\cos^{2/s}(2\phi/\phi_0),
\end{eqnarray}
where $\phi_0=\frac{\sqrt{3(2s-1)}}{s}$ with $s=1,3,5...$, and $G(z)$, ${\cal K}(\phi)$ are defined respectively as
\begin{equation}
G(z) \equiv 1+a~z~{}_2F_1
      \left(
            \frac{1}{2s},\frac{1}{s},1+\frac{1}{2s},-(\lambda z)^{2s}
      \right),\label{G}
\end{equation}
\begin{equation}
{\cal K}(\phi)\equiv
   \lambda+a\tan^{1/s}\big({\phi}/{\phi_0}\big)~
      {}_2F_1\left(\frac{1}{2s},\frac{1}{s},1+\frac{1}{2s}
              ,-\tan^{2}\big(\frac{\phi}{\phi_0}\big)\right)
\end{equation}
with ${}_2F_1$ the hypergeometric function. This is a three-parameter ($\lambda, s, a$) family of asymmetric thick brane. And $a$ is called asymmetric factor satisfying
\begin{equation}
0\leq a<\frac{2~s~\lambda~\Gamma(1/s)}{\Gamma(1/2s)^2}.
\end{equation}
As $a=0$, we will get a symmetric solution. And for $s=1$ and $s>1$, the domain wall is a single brane and double one, respectively. Thus, these branes have very rich inner structure. We will mainly investigate the effect of the asymmetry on resonance.

Therefore, we consider KR field $B_{\mu\nu}$ in this space-time, and regard the field as a small perturbation around the background. The action of the field is given by
\begin{equation}
S_{KR}=\int d^5x\sqrt{-g}~\big[
                                f(\phi)H_{MNL}H^{MNL}
                         \big],\label{action}
\end{equation}
where $H_{MNL}=\partial_{[M}B_{{NL}]}$ is the field strength, and $f(\phi)$ is the coupling to the background scalar field. The reason we introduce the coupling is that the free KR field cannot be localized on the brane~\cite{Cruz2009,Tahim2009}.

From the action (\ref{action}) and the metric (\ref{metric}), the equations of motion for the KR field read
\begin{eqnarray}
f(\phi)\partial_{\mu}[\sqrt{-g}H^{\mu\nu\lambda}]
+\partial_z[\sqrt{-g}f(\phi)H^{z\nu\lambda}]=0,\label{motion}\\
f(\phi)\partial_{\mu}[\sqrt{-g}H^{\mu\nu z}]=0.
\end{eqnarray}

In the following, we investigate the KK modes of the KR field, which not only can produce the four-dimensional KR fields, but also carry the information of the extra dimension.  First, we choose a gauge $B_{\mu z}=0$, and decompose the field as
\begin{equation}
B^{\nu\lambda}(x,z)=\sum_nb^{\nu\lambda}_{(n)}(x)~\chi_n(z)~e^{pA(z)}\label{decompsition1}
\end{equation}
with $p$ a coupling constant. Then the field strength becomes
\begin{eqnarray}
H_{\mu\nu\lambda}&=&\sum_n e^{4A(z)}~h_{\mu\nu\lambda(n)}(x)\chi_n(z)~e^{pA(z)},
\label{decompsition1}
\end{eqnarray}
where $h_{\mu\nu\lambda(n)}(x)=\partial_{[\mu} b_{\nu\lambda](n)}(x)$ is the field strength on the brane. Here, we consider the coupling $f(\phi(z))=e^{(-7-2p)A(z)}$. Then, by substituting (\ref{decompsition1}) into Eq. (\ref{motion}), we obtain the Schr\"{o}dinger-like equation for the KK modes $\chi_n(z)$:
\begin{equation}
\bigg[-\frac{d^2}{dz^2}+V(z)\bigg]\chi_n(z)=m_n^2\chi_n(z),\label{SLE}
\end{equation}
where $m_n$ are the masses of the four-dimensional KR fields, and the effective potential $V(z)$ takes the form
\begin{equation}
V(z)=(4+p)^2A'^{2}(z)-(4+p)A''(z).\label{potential}
\end{equation}

Because the Schr\"{o}dinger-like equation (\ref{SLE}) can be written in supersymmetric quantum mechanics scenario as follows:
\begin{eqnarray}
Q^\dag Q~\chi_n(z)=\bigg[
                       -\frac{d}{dz}+(4+p)A'(z)
                   \bigg]
                   \bigg[
                          \frac{d}{dz}+(4+p)A'(z)
                   \bigg]\chi_n(z)
                 =m_n^2\chi_n(z),\label{superpotential}
\end{eqnarray}
we can exclude the existence of tachyonic modes if the KK modes are regular at the boundaries, and this is an essential condition to keep the stability of the model.

On the other hand, we need the orthonormality conditions
\begin{eqnarray}
\int^\infty_{-\infty} dz~\chi_n(z)\chi_l(z)=\delta_{nl},
\label{orthonormalityconditions}
\end{eqnarray}
in order to obtain the effective action for the four-dimensional KR fields:
\begin{eqnarray}
S_{KR}=\sum_n\int d^4x
        \big[
               h_{\mu\nu\lambda(n)}(x)~h^{\mu\nu\lambda}_{(n)}(x)
               +3~m^2_n~b_{\nu\lambda(n)}(x)~b^{\nu\lambda}_{(n)}(x)
        \big].
\end{eqnarray}
Thus, we can use the orthonormality conditions (\ref{orthonormalityconditions}) to judge whether the KK modes of the KR field can be localized on the symmetric and asymmetric branes.

\subsection{The symmetric brane case}\label{scheme2}

\begin{figure*}[htb]
\begin{center}
\includegraphics[width=6cm]{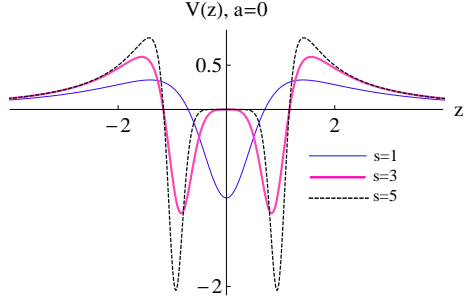}
\end{center}
\vskip -4mm \caption{The shape of the symmetric potential $V^s(z)$ with $\lambda=1$ and $p=-5$.}\label{figVS}
\end{figure*}

Firstly, we investigate the case of the symmetric thick brane ($a=0$). The effective potential (\ref{potential}) is symmetric, and takes the following form
\begin{eqnarray}
V^s(z)&=&(4+p)(\lambda z)^{2s}
              \frac{(3+p)(\lambda z)^{2s}+(2s-1)}
                   {z^2\big[1+(\lambda z)^{2s}\big]^2}.
                     \label{symmetricpotential}
\end{eqnarray}
When $z\rightarrow\infty$, $1+(\lambda z)^{2s}\rightarrow(\lambda z)^{2s}$, so we have
\begin{eqnarray}
V^s(z\rightarrow\infty)\rightarrow\frac{(4+p)(3+p)}{z^2}
 +\frac{(4+p)(2s-1)}{z^2(\lambda z)^{2s}}\rightarrow 0.
\end{eqnarray}
And when $z\rightarrow 0$, $1+(\lambda z)^{2s}\rightarrow 1$, we find
\begin{eqnarray}
V^s(z\rightarrow 0)&\rightarrow&\frac{(4+p)(3+p)(\lambda z)^{4s}}{z^2}
 +\frac{(4+p)(2s-1)(\lambda z)^{2s}}{z^2}\nonumber\\
 &\rightarrow&\left\{
  \begin{array}{ll}
    \lambda^2(4+p) & ~~~~~~~~~~~\text{for}~~~~~ s=1 \\
    0 & ~~~~~~~~~~~\text{for}~~~~~ s> 1
  \end{array}\right..~~~~~
\end{eqnarray}
It can be seen that $V^s(z)$ is a volcano potential with a single well for $s=1$ and two ones for $s\geq3$. And the distance between the two minima of $V^s(z)$ for $s\geq3$ increases with $s$. The parameter $p$ mainly effects the depth of the potential, and with the increase of $p$ the potential well is deeper. The width of the potential is controlled by the parameter $\lambda$; when $\lambda$ increases, the potential well becomes narrower. We only plot the shapes of $V^s(z)$ in Fig.~\ref{figVS} for different values of $s$.

For this type of potential, there are a zero mode and continuous massive modes. Some resonant states may appear. For the zero mode ($m_0^2=0$), the wave function is
\begin{equation}
\chi_0^s(z)\propto
            \bigg(
                  1+(\lambda z)^{2s}
            \bigg)^{\frac{4+p}{2s}}.
\end{equation}
To see whether the zero mode can be localized on the brane, we need to check the normalization condition (\ref{orthonormalityconditions}) for $\chi_0^s(z)$. Because
\begin{eqnarray}
I^{s}=\int^\infty_{-\infty} (\chi_0^s)^2 dz \propto
 \int^\infty_{-\infty}
  \bigg(
         1+(\lambda z)^{2s}
  \bigg)^{\frac{4+p}{s}} dz
  \rightarrow\int^\infty_{-\infty} (\lambda z)^{2(4+p)}~dz~~~~\text{for}~~~ z\rightarrow \infty,
  \label{Is}
\end{eqnarray}
which is finite for $p<-9/2$. So, the zero mode can be localized on the symmetric thick brane for $p<-9/2$.

In order to find the resonances for this symmetric potential, we can use the method given in Refs.~\cite{Liu2009a,Liu2009c}, where the authors defined a variable, called the relative probability in a box with borders at $\pm z_{max}$:
\begin{equation}
P=\frac{\int_{-z_b}^{z_b}|\chi_n(z)|^2 dz}
            {\int_{z_{max}}^{-z_{max}}|\chi_n(z)|^2 dz},
\end{equation}
and the large relative probability for massive KK modes within a range $-z_b<z<z_b$ indicates the existence of resonances. We note here that this method is the extended version of the one used in Ref. \cite{Almeida2009}, which is only effective for the even KK modes. On the other hand, because the potential is symmetric, the wave functions are either even-parity or odd-parity. So one can give two additional boundary conditions to solve the differential equation  (\ref{SLE}) numerically:
\begin{eqnarray}
\label{incondition}
  \chi_n(0)&=&0, ~~\chi_n'(0)=c_0,~~~~~~\text{for}~~~~\text{odd-parity},\\
  \chi_n(0)&=&c_1, ~~\chi_n'(0)=0,~~~~~~\text{for}~~~~\text{even-parity}.
\end{eqnarray}

\begin{figure*}[htb]
\begin{center}
\includegraphics[width=6cm,height=4.5cm]{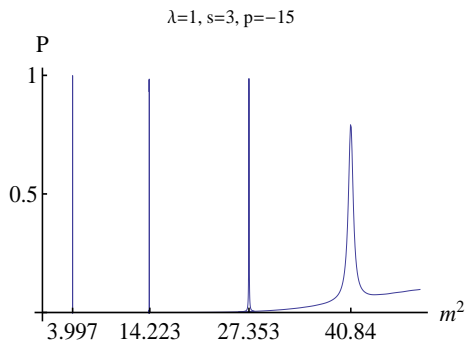}
\end{center}
\vskip -4mm \caption{The shape of the relative probability $P$ (which is a function of $m^2$) of the massive KK modes for the symmetric case with $\lambda=1,~s=3,~p=-15$. }
\label{figReS}
\end{figure*}

Here, we give an example. We set the parameters as $a=0$, $\lambda=3$, $s=1$, $p=-15$ and $z_{max}=20z_0$, and find some resonances. The shape of the relative probability is plotted in Fig.~\ref{figReS}, where the masses of resonances are marked, i.e., $m_n^2=3.997,~14.223,~27.353,~40.84$.

\subsection{The asymmetric brane case}\label{sec22}

In this subsection, we consider the case of asymmetric thick brane $(a\neq0)$. The asymmetric effective potential $V^a(z)$ reads
\begin{eqnarray}
V^a(z)&=&(4+p)(\lambda z)^{2s}
                \frac{(3+p)(\lambda z)^{2s}+(2s-1)}
                     {z^2\big[1+(\lambda z)^{2s}\big]^2}\nonumber\\
       &&+(4+p)(3+p)\frac{G^{'2}}{G^2}
     +(4+p)\frac{G^{''}}{G}
       +\frac{2(4+p)^2(\lambda z)^2G'}{z[1+(\lambda z)^{2s}]G}.
\end{eqnarray}
It is also a volcano-like one. So there are also a zero mode and continuous massive modes, and some resonances may exist.

With the warp factor (\ref{warpfactor}), the zero mode wave function is given by
\begin{equation}
\chi_0^a(z)\propto[1+(\lambda z)^{2s}]^{\frac{4+p}{2s}}G^{4+p}(z).
\end{equation}
Then we consider the normalization condition for $\chi_0^a(z)$:
\begin{eqnarray}
 I^a=\int_{-\infty}^\infty [~\chi^a_0(z)~]^{2}~dz
   \propto \int_{-\infty}^\infty [~\chi^s_0(z)~]^{2}~G^{2(4+p)}(z)~dz.
\end{eqnarray}
Noting that, when $z\rightarrow\pm\infty$, $G(z) \rightarrow 1\pm\frac{a~\Gamma(\frac{1}{2s})~
\Gamma(\frac{1}{2s}+1)}{\lambda~\Gamma(\frac{1}{s})}$, which is an finite constant, we get the conclusion that
the condition for localizing the zero mode in this asymmetric brane is the same with that on the symmetric case, i.e., $p<-9/2$.

Next, we turn to analyze resonances of the KR field on the asymmetric brane. Because the effective potential is asymmetric, the corresponding wave functions have no definite parity. When solving the second-order differential equation (\ref{SLE}), we cannot use the initial conditions (\ref{incondition}). So the method for finding the resonances in Sec. \ref{scheme2} is not suitable for this asymmetric case. Here, we can use the transfer matrix method. This method has been used to investigate the resonances of matter fields in the RS model in Refs.~\cite{Landim2011,Landim2012,Alencar2012}. Here we use this method to solve the resonances of KR fields on this asymmetric brane.

In this method, a transmission coefficient $T$ is introduced:
\begin{equation}
T=\frac{k_1}{k_N}\frac{1}{\mid M_{22}\mid^2},
\end{equation}
where $N$ is the number of the transfer coefficient matrixes, and $M$ is the product of the $N$ transfer coefficient matrixes with dimension $2\times 2$. The relation between $k_1$ and $k_N$ is
\begin{eqnarray}
\frac{k_1}{k_N}=\sqrt{\frac{\mid V_1-E_n\mid}{\mid V_N-E_n\mid}}
\end{eqnarray}
with $E_n=m_n^2$.

The transmission coefficient $T$ gives a clear interpretation of what happens to a plane wave interacting with the branes, no matter whether the brane is symmetric. For a given mass, we will get a $T(m^2)$. If the transmission coefficient has a peak, then it means that the amplitude of the wave function corresponding to the peak has relative larger value on the brane, namely, we will have a higher probability to find this KK mode on the brane; hence we will call this KK mode as a resonant state.

\begin{figure*}[htb]
\begin{center}
\includegraphics[width=6cm]{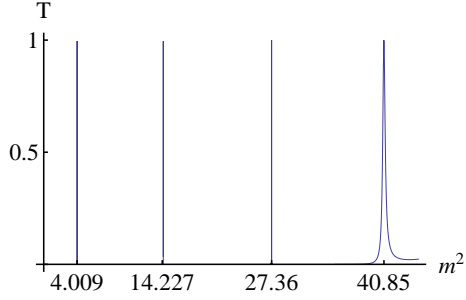}
\end{center}
\vskip -4mm \caption{The transmission coefficient $T$ for the KK modes of the KR field with $p=-15,~\lambda=1,~s=1,~a=0$}\label{figTsy}
\end{figure*}

Thus, we try to use this transfer matrix method to find the resonances in both the symmetric and asymmetric branes, and compare the results obtained by the transfer matrix method with that by the method in Sec.~\ref{scheme2} for the symmetric case. Here, we set $N=10^4+1$, and $V_1=V(-z_{max})$, $V_N=V(z_{max})$ with $z_{max}=20$.

We first plot the shape of the transmission coefficient as a function of $m^2$ for the symmetric case in Fig.~\ref{figTsy} with $a=0,~\lambda=1,~s=3,~p=-15$. Comparing Fig.~\ref{figTsy} with Fig.~\ref{figReS}, we find that, for the symmetric case, the resonances obtained with the transfer matrix method is almost the same with the result obtained with the relative probability method used in Sec.~\ref{scheme2}. In fact, the relative probability $P$ and the transmission coefficient $T$ have the same physical meaning, i.e., they indicate the interaction between a plane wave function and the brane. Under the same condition, the resonances obtained by the two methods should be the same.

\begin{figure*}[htb]
\begin{center}
\subfigure[$p=-30,\lambda=1,s=1$]{\label{figTa1}
\includegraphics[width=6cm]{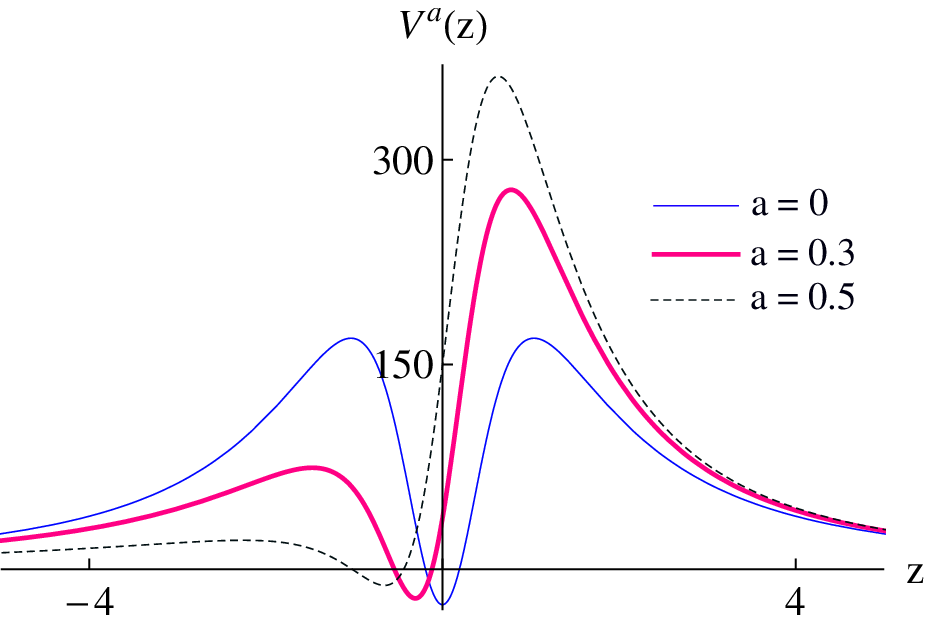}~~~~~~~~~
\includegraphics[width=6cm]{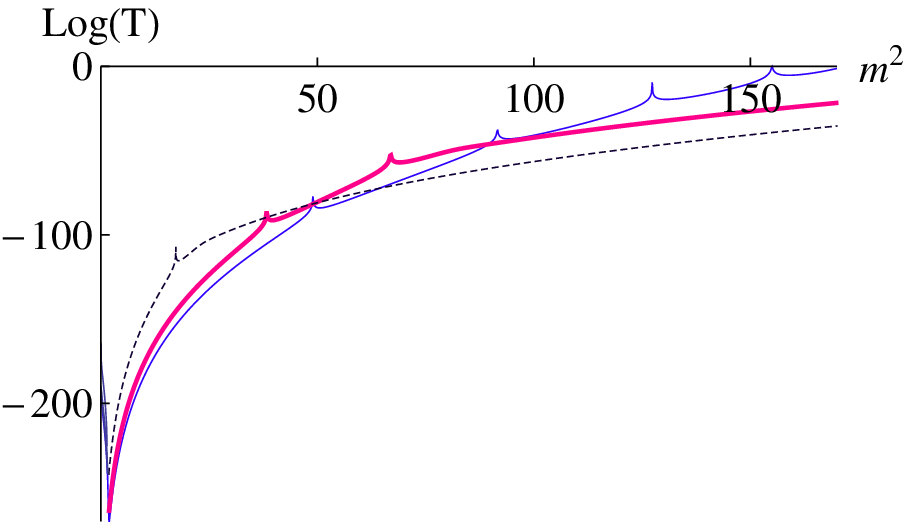}}\\
\subfigure[$p=-30,\lambda=1,s=3$]{\label{figTa2}
\includegraphics[width=6cm]{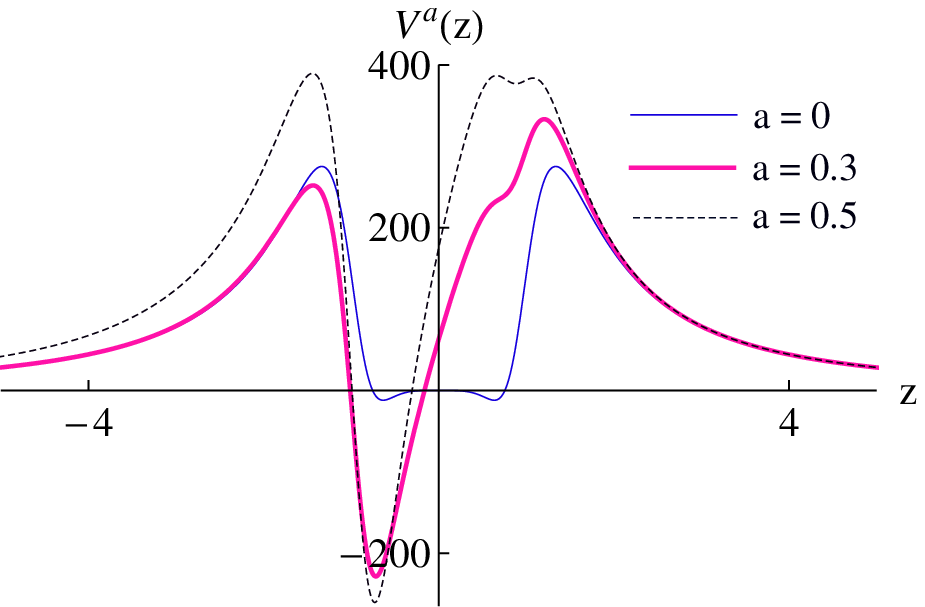}~~~~~~~~~
\includegraphics[width=6cm]{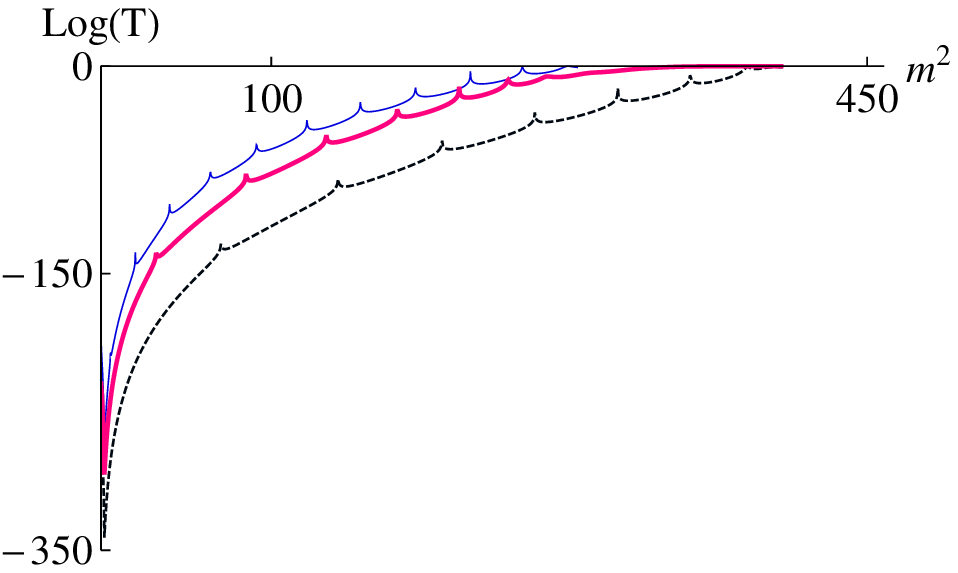}}\\
\subfigure[$p=-30,\lambda=1,s=5$]{\label{figTa3}
\includegraphics[width=6cm]{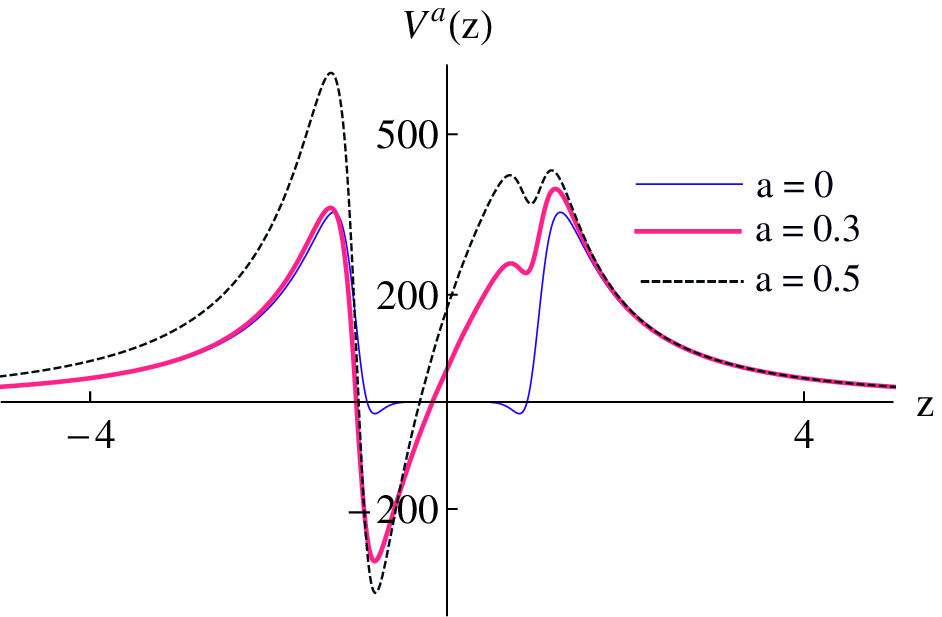}~~~~~~~~~
\includegraphics[width=6cm]{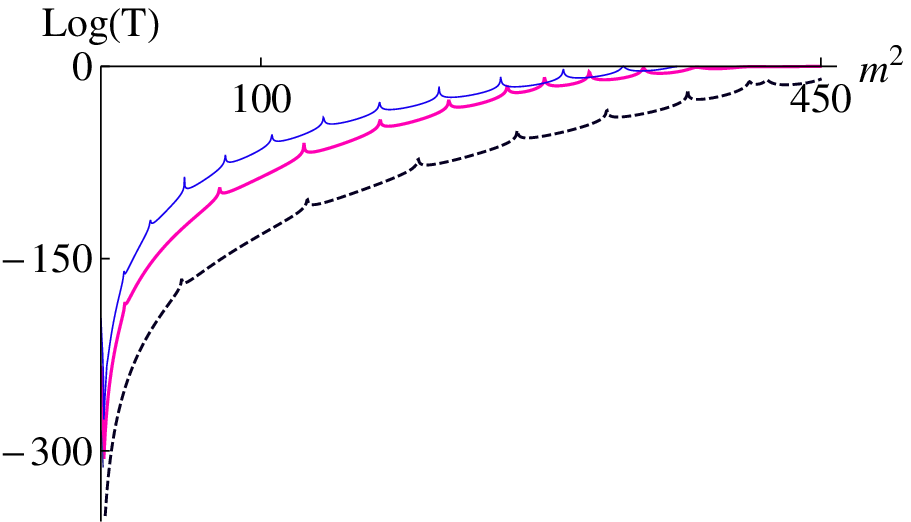}}
\end{center}
\vskip -4mm \caption{The logarithm for the KK modes of the KR field on the symmetric and asymmetric thick branes with different parameters.}\label{figTa}
\end{figure*}

In order to find the effect of the asymmetry factor $a$ on the resonances, we plot the shapes of the effective potential $V^a(z)$ with different values of $a$ (including $a=0$) in Figs.~\ref{figTa1}, \ref{figTa2} and \ref{figTa3}, for $s=1$, $s=3$ and $s=5$, respectively. The corresponding logarithm of the transmission coefficient as a function of $m^2$ is also given in these figures. The masses of the resonances for these cases are listed in Tab.~\ref{tableA}.

\begin{table}
\centering
\begin{tabular}{|c|c|c||c|c|c||c|c|c|}\hline
\multicolumn{3}{|c||}{$a=0~(m_n^{s~2})$}&
\multicolumn{3}{|c||}{$a=0.3~(m_n^{a~2})$}&\multicolumn{3}{|c|}{$a=0.5~(m_n^{a~2})$}
\\\hline
\multicolumn{1}{|c|}{$s=1$}&\multicolumn{1}{|c|}{$s=\!3$}&
\multicolumn{1}{|c||}{$s=\!5$}&
\multicolumn{1}{|c|}{$s=\!1$}&\multicolumn{1}{|c|}{$s=\!3$}&
\multicolumn{1}{|c||}{$s=\!5$}&
\multicolumn{1}{|c|}{$s=\!1$}&
\multicolumn{1}{|c|}{$s=\!3$}&\multicolumn{1}{|c|}{$s=\!5$}
\\\hline
$49.08$&$5.64$&$14.27$&
$38.24$&$32.16$&$14.99$&$17.28$&$70.15$&$50.40$
\\\hline
$91.74$&$20.07$&$31.03$&
$66.8$&$85.17$&$74.13$&-&$139.3$&$128.6$
\\\hline
$127.4$&$40.27$&$52.32$&
-&$132.3$&$126.9$&-&$200.6$&$198.1$
\\\hline
-&$64.14$&$77.68$&
-&$174$&$174.7$&-&$254.9$&$259.8$
\\\hline
-&$91.16$&$107.1$&
-&$210.3$&$217.3$&-&$303.4$&$316.2$
\\\hline
-&$121.1$&$138.8$&
-&$239.1$&$254.1$&-&$346.2$&$366.8$
\\\hline
-&$152.3$&$174.2$&
-&-&$277.3$&-&$379.5$&$405$
\\\hline
-&$184.8$&$211.3$&
-&-&$305.2$&-&-&$416.1$
\\\hline
-&$217.1$&$250$&
-&-&$339$&-&-&-
\\\hline
-&$247.5$&$289.2$&
-&-&-&-&-&-
\\\hline
-&-&$326.4$&
-&-&-&-&-&-\\\hline
\end{tabular}
\caption{The masses of resonances for the KK modes on symmetric and asymmetric thick branes. The parameters are set to $p=-30$, $\lambda=1$.}
\label{tableA}
\end{table}

From Fig.~\ref{figTa} and Tab.~\ref{tableA}, we find that:
\begin{itemize}
  \item{For $s=1$ and $s=5$, comparing with the symmetric case $a=0$, there are less number of the resonances for the asymmetric case $a\neq0$. And with the increase of $a$, the number will decrease.}
  \item{For $s=3$, the number of the resonances is the most for the symmetric case. While it does not monotonically decrease with $a$ for the asymmetric case.}
\end{itemize}
And it seems that the $a$ does not decide the number of resonances. While the degree of the asymmetry plays the role. When the degree of the asymmetry increases, the number of resonances is less. For $s=3$, the degree of the asymmetry for the potential seems deeper for $a=0.3$ comparing with $a=0.5$, so the number of the resonances is less for $a=0.3$.

\section{Conclusion}\label{secConclusion}

In this paper, we investigate the localization of KR field on symmetric and asymmetric thick branes. There are two important parameters in the brane model, i.e., $s$ and $a$. The former decides the splitting of the brane, and the latter leads to the asymmetry of the brane, so it is called as asymmetric factor. For $a=0$, the brane is symmetric, and for $a\neq0$ the brane is asymmetric. We mainly investigate the effect of the asymmetry on the localization.

We consider the coupling between the KR field and the background scalar field ($f(\phi(z))=e^{(-7-2p)A(z)}$), and find that the KK modes of the KR field satisfy a Schr\"{o}dinger-like equation. Because the effective potential is a volcano-like one for both symmetric and asymmetric cases, there are always a zero mode and continuous massive modes, and some resonances may exist. The condition for localizing the zero mode on both the symmetric and asymmetric branes is the same, i.e., $p<-9/2$.

In order to find the resonances, we first use the relative probability method for the symmetric case, but it is not suitable for the asymmetric case. Then we use the transfer matrix method. This method can be used in both cases. And for the symmetric case, we obtain the same result for the two methods. This may be because both the relative probability $P$ (in the relative probability method) and the transmission coefficient $T$ (in the transfer matrix method) indicate the interaction of a wave function with the brane.

The number of the resonances for the asymmetric case is decided by the degree of the asymmetry, but not completely the asymmetric factor $a$. The number of the resonances decreases with the degree of the asymmetry.

\section*{Acknowledgement}

We especially thank Prof. Y. X. Liu and Prof. R. R. Landim for sharing some ideas about resonance state and the codes. We also thank Xiao-Long Du, Xiang-Nan Zhou and Ke Yang for helpful discussions. This work was supported in part by the National Natural Science Foundation of China (No. 11075065 and No. 10905027), the Huo Ying-Dong Education Foundation of
Chinese Ministry of Education (No. 121106) and the Fundamental Research Funds for the Central Universities.





\end{document}